\documentclass{webofc}
\usepackage[varg]{txfonts}   % Web of Conferences font

\newcommand{\GeVc}{\ensuremath{\mbox{Ge\kern-0.1em V}\!/\!c}\xspace}

\newcommand{\AGeV}{\ensuremath{A\,\mbox{Ge\kern-0.1em V}}\xspace}
\newcommand{\AGeVc}{\ensuremath{A\,\mbox{Ge\kern-0.1em V}\!/\!c}\xspace}
\newcommand{\dd}{\ensuremath{{\textrm d}}\xspace}

\begin{document}
\title{Results on system size dependence of strangeness production in the CERN SPS energy range from NA61/SHINE}

\author{\firstname{Piotr} \lastname{Podlaski}\inst{1}\fnsep\thanks{\email{piotr.podlaski@cern.ch}} for the NA61/SHINE Collaboration
        % etc.
}

\institute{Faculty of Physics, University of Warsaw, Warsaw, Poland
          }

\abstract{%
  NA61/SHINE is a multipurpose fixed-target facility at the CERN Super Proton Synchrotron. The main goals of the NA61/SHINE strong interactions program are to discover the critical point of strongly interacting matter and study the properties of the onset of deconfinement. To reach these goals, hadron production measurements are performed in the form of a two-dimensional scan by varying collision energy and system size. The Collaboration has recently finished data acquisition for its original program on strong interactions, accumulating broad data samples on hadron production in various systems in the SPS energy range.\\
In this contribution, the NA61/SHINE results on identified charged kaon and pion production in p+p, Be+Be and Ar+Sc collisions at the SPS energy range ($\sqrt{s_{NN}}$=5.1--17.3 GeV) are presented. The NA61/SHINE measurements of small and intermediate-mass ion collisions establish an interesting system size dependence, showing a rapid change of hadron production properties that starts when moving from Be+Be to Ar+Sc system. In particular, Ar+Sc is the smallest system for which a significant enhancement of $K^+/\pi^+$ ratio with respect to $p$+$p$ collisions is observed. Obtained energy and system size dependence of the measured charged hadron multiplicities are compared with available world data and various theoretical models.
}
\maketitle
\section{Introduction}
\label{intro}
NA61/SHINE is a large acceptance hadron spectrometer located in the CERN's North Area \cite{fac_paper}. Eight large volume Time Projection Chambers (TPC), accompanied by the Time of Flight detectors (TOF) provide tracking and identification of produced particles. The Projectile spectator Detector (PSD), a precise zero-degree hadron calorimeter, measures the energy of projectile spectators, which can be related to the centrality of the collision.\\
NA61/SHINE performed a unique, two-dimensional scan in system size ($p$+$p$, $p$+Pb, Be+Be, Ar+Sc, Xe+La, Pb+Pb) and momentum of the beam (13$A$ -- 150\AGeVc). NA61/SHINE physics program focuses on studies of the phase diagram of strongly interacting matter and searches for the critical point.
\section{Studies of the onset of deconfinement}
\label{sec-ood}
A first-order phase transition between Quark-Gluon Plasma and hadron gas was predicted by the Statistical Model of The Early Stage (SMES) \cite{smes1,smes2} to be located in the CERN SPS energy range. 
\begin{figure}[ht]
\centering
\includegraphics[width=0.36\textwidth]{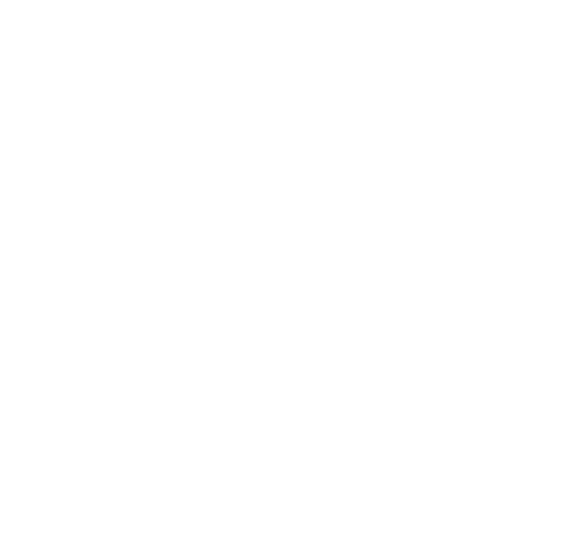}~~~~
\includegraphics[width=0.36\textwidth]{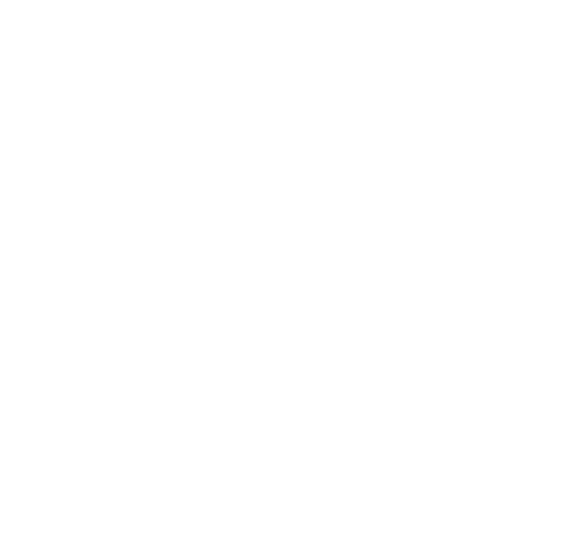}
\vspace{-0.45cm}

\caption{Energy dependence of the inverse slope parameter of $K^+$ (\textit{left}) and energy dependence of the $K^+/\pi^+$ particle yield ratio in mid-rapidity (\textit{right}) for $p$+$p$, Be+Be, Ar+Sc and Pb+Pb collisions \cite{pp_paper,star_pp, compilation_pp,alice_1,alice_3, Be_paper, star_bes,star_bes_2,phenix,phenix_2,brhams,AGS_scan,NA49_Pb,na49_pb_2,na49_pb_3,alice_2,HADES:2017jgz}.}
\label{fig-step-horn}
\end{figure}
Left panel of Fig. \ref{fig-step-horn} shows the inverse slope parameter of the positive kaon $m_T$ spectra at midrapidity for different colliding systems as a function of collision energy. Within SMES the characteristic plateau in the energy dependence of the inverse slope parameter for heavy ion collisions (Pb+Pb, Au+Au) can be attributed to the existence of the mixed phase of QGP and hadron gas. The NA61/SHINE results on light and intermediate systems ($p$+$p$, Be+Be, Ar+Sc) exhibit a qualitatively similar behavior, with the plateau value growing with the system size. The most prominent signature of the phase transition, predicted within SMES, is a rapid, non-monotonic change of the $K^+/\pi^+$ ratio as a function of collision energy, the horn. The right panel of Fig. \ref{fig-step-horn} presents a compilation of results on positive kaon to positive pion multiplicity ratio at midrapidity. The horn structure was not observed in data on central Ar+Sc collisions. Moreover, Ar+Sc results on $K^+/\pi^+$ ratio at low SPS energies are located between light ($p$+$p$, Be+Be) and heavy (Pb+Pb, Au+Au) systems, while for higher collision energies the $K^+/\pi^+$ ratio is following the heavy systems.

\begin{figure}[b]
\centering
\includegraphics[width=0.32\textwidth]{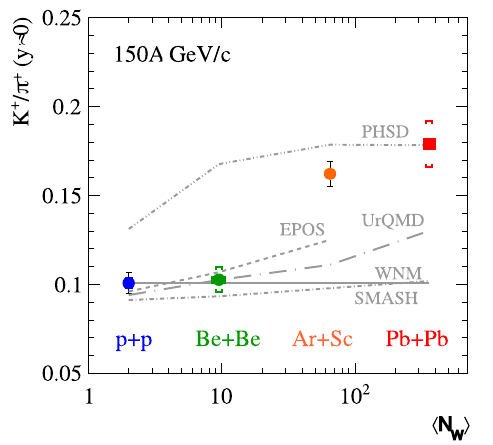}
\includegraphics[width=0.32\textwidth]{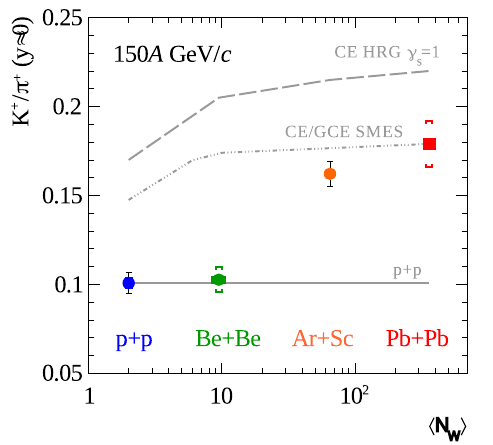}
\includegraphics[width=0.32\textwidth]{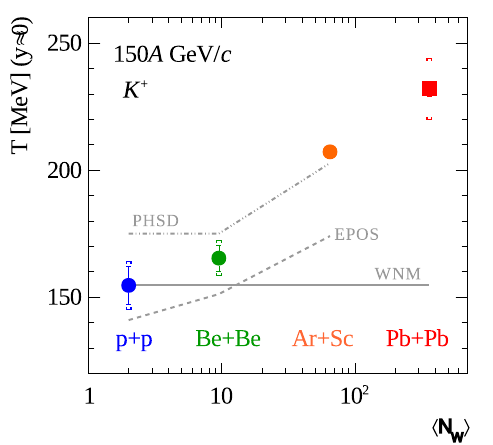}
\vspace{-0.4cm}
\caption{System size dependence of the $K^+/\pi^+$ ratio at midrapidity measured at 150\AGeVc compared with dynamical (\textit{left}) and statistical (\textit{center}) models. \textit{Right}: System size dependence of the inverse slope parameter $T$ of $K^+$ at the same collision energy \cite{pp_paper,Be_paper,PHSD,urqmd,smes2,NA49_Pb}.}
\label{fig-syst-size}
\end{figure}

Figure \ref{fig-syst-size} presents $K^+/\pi^+$ multiplicity ratio and inverse slope parameter of the $K^+$ transverse mass spectra as a function of the system size for the top SPS energy (150\AGeVc beam momentum). System size is quantified by the mean number of wounded nucleons in a collision $\langle W \rangle$. Dynamical models without phase transition (EPOS, PHSD, UrQMD) successfully describe the $K^+/\pi^+$ ratio for light systems ($p$+$p$ and Be+Be) but fail for heavier ones (Ar+Sc, Pb+Pb). On the other hand PHSD, the model with phase transition, reproduces the data for heavy systems but overestimates the $K^+/\pi^+$ ratio for lighter ones. Both statistical models taken into consideration tend to overestimate the data, regarding if they include the phase transition (SMES) or not (HRG). Both considered quantities ($K^+/\pi^+$ ratio and inverse slope parameter $T$) show qualitatively similar system size dependence, which cannot be fully described by any of the statistical or dynamical models used for comparison.

\section{Strangeness production in $p$+$p$ interactions}
\label{sec-strange_protons}
\begin{figure}[hb]
\centering
\includegraphics[width=0.8\textwidth]{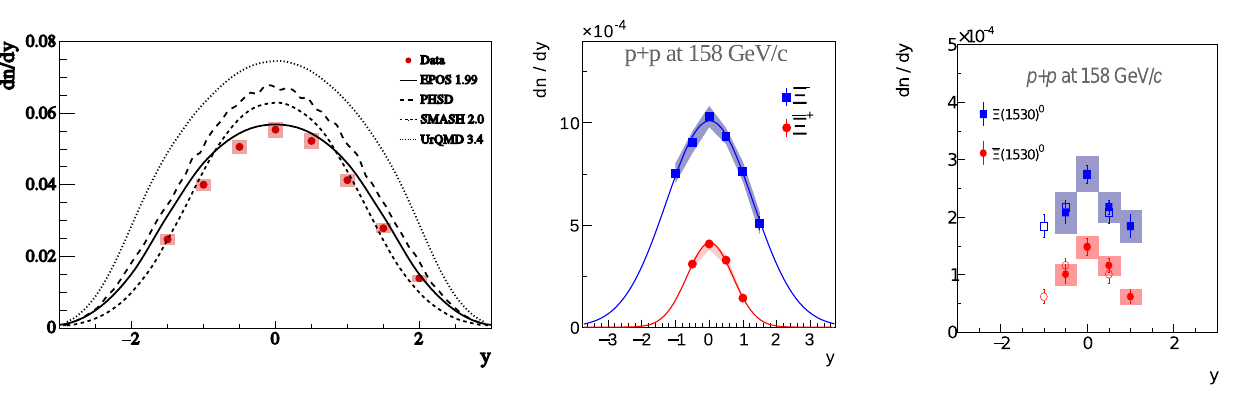}
\vspace{-0.45cm}
\caption{Rapidity spectra of: $K_S^0$ (\textit{left}), $\Xi^+$ and $\bar{\Xi}^-$ (\textit{center}), $\Xi^0(1530)$ and $\bar{\Xi}^0(1530)$ (\textit{right}) in $p$+$p$ interactions at 158 \GeVc \cite{k0s_paper,xi_paper,xi_paper_err,xi_1530_paper}.}
\label{fig-strange-spectra}
\end{figure}

The NA61/SHINE has accumulated almost 30 million $p$+$p$ collisions at 158 \GeVc beam momentum, allowing for precise studies of the strangeness production mechanisms in elementary collisions, as well as obtaining references to A+A systems. The left panel of Fig. \ref{fig-strange-spectra} shows the measured rapidity spectrum of $K^0_S$ mesons produced in $p$+$p$ interactions at 158 \GeVc. The measured mean multiplicity of $\langle K_S^0 \rangle$ is $0.162\pm0.001\pm0.011$. This result, accompanied by the energy scan of $K_S^0$ production in $p$+$p$ interactions, will serve as reference measurements for future analyses of collisions of heavier nuclei.

Unique, high statistics $p$+$p$ dataset allows for studies of multi-strange baryon production by the NA61/SHINE. Center and right panels of Fig. \ref{fig-strange-spectra} presents rapidity spectra of $\Xi^+$, $\bar{\Xi}^-$, $\Xi^0(1530)$ and $\bar{\Xi}^0(1530)$ measured in $p$+$p$ collisions at 158 \GeVc. 
The production of anti-$\Xi$ hyperons is suppressed compared to the production of $\Xi$ hyperons, namely the multiplicity ratios are the following:
%Strong suppression of $\bar{\Xi}^-$ production was observed, provided that the 
$\overline{\Xi}^+/\Xi^-$ ratio is $0.24 \pm 0.01 \pm 0.05$. 
%Production of $\bar{\Xi}^0(1530)$ was slightly less suppressed, measured corresponding multiplicity ratio
and $\overline{\Xi}^0(1530)/\Xi^0(1530)$ ratio is $0.40 \pm 0.03 \pm 0.05$. Noteworthy, $\Xi^0(1530)$ measurement by the NA61/SHINE is the only one available in the SPS energy range.

\begin{figure}[ht]
\centering
\includegraphics[width=0.65\textwidth, trim= 0 7 0 5, clip]{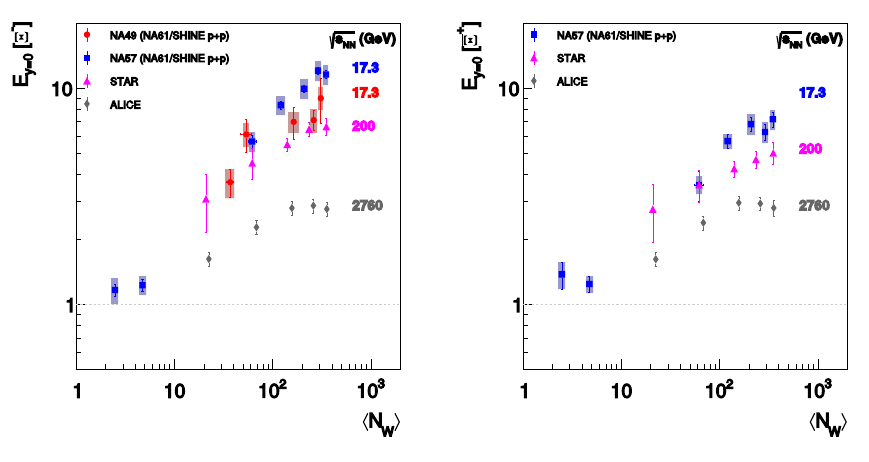}
\vspace{-0.4cm}
\caption{Strangeness enhancement factors calculated using $\Xi^+$ and $\bar{\Xi}^-$ yields measured by NA61/SHINE \cite{xi_paper,xi_paper_err}. }
\label{fig-enh}
\end{figure}

The $\Xi$ multiplicities were used to calculate strangeness enhancement factors, $E$. The $E$ is defined as follows:
\begin{equation}
    E=\frac{2}{\langle W\rangle} \frac{\dd n/\dd y (A+A)}{\dd n / \dd y (p+p)}
\end{equation}
where the midrapidity density of $\Xi$ in A+A collision, scaled by the mean number of wounded nucleons, is divided by the corresponding density measured in inelastic $p$+$p$ collisions ($W=2$ for protons). Fig. \ref{fig-enh} shows enhancement factors for $\Xi^+$, $\bar{\Xi}^-$, recalculated using the latest NA61/SHINE results, as a function of the mean number of wounded nucleons. The NA61/SHINE provided a new reference for the calculation of strangeness enhancement factors in A+A collisions.

\newpage
{
\footnotesize
\textit{Acknowledgments:} This work was supported by the Polish Minister of Education and Science (contract No. 2021/WK/10).
}

\bibliography{bibliography}

\end{document}